\documentclass[a4paper]{jpconf}
\usepackage{graphicx}
\begin{document}
\title{Improvements of the ALICE HLT data transport framework for LHC Run~2}

\author{David Rohr$^{1,2}$, Mikolaj Krzwicki$^1$, Heiko Engel$^3$, Johannes Lehrbach$^1$, Volker Lindenstruth$^{1,3}$ for the ALICE Collaboration}

\address{$^1$ Frankfurt Institute for Advanced Studies, Ruth-Moufang-Str. 1, 60438 Frankfurt, Germany\\
         $^2$ CERN, 385 Route de Meyrin, 1217 Meyrin, Switzerland\\
         $^3$ Institut f\"ur Informatik, Johann Wolfgang Goethe-Universit\"at Frankfurt, Robert-Mayer-Str. 11-15, 60629 Frankfurt, Germany\\}

\ead{drohr@compeng.uni-frankfurt.de}

\begin{abstract}
The ALICE HLT uses a data transport framework based on the publisher-subscriber message principle, which transparently handles the communication between processing components over the network and between processing components on the same node via shared memory with a zero copy approach.
We present an analysis of the performance in terms of maximum achievable data rates and event rates as well as processing capabilities during Run~1 and Run~2.
Based on this analysis, we present new optimizations we have developed for ALICE in Run~2.
These include support for asynchronous transport via Zero-MQ which enables loops in the reconstruction chain graph and which is used to ship QA histograms to DQM.
We have added asynchronous processing capabilities in order to support long-running tasks besides the event-synchronous reconstruction tasks in normal HLT operation.
These asynchronous components run in an isolated process such that the HLT as a whole is resilient even to fatal errors in these asynchronous components.
In this way, we can ensure that new developments cannot break data taking.
On top of that, we have tuned the processing chain to cope with the higher event and data rates expected from the new TPC readout electronics (RCU2) and we have improved the configuration procedure and the startup time in order to increase the time where ALICE can take physics data.
We analyze the maximum achievable data processing rates taking into account processing capabilities of CPUs and GPUs, buffer sizes, network bandwidth, the incoming links from the detectors, and the outgoing links to data acquisition.
\end{abstract}

\section{Introduction}
\begin{figure}[htb]
\begin{minipage}{16.5pc}
\includegraphics[width=16.5pc]{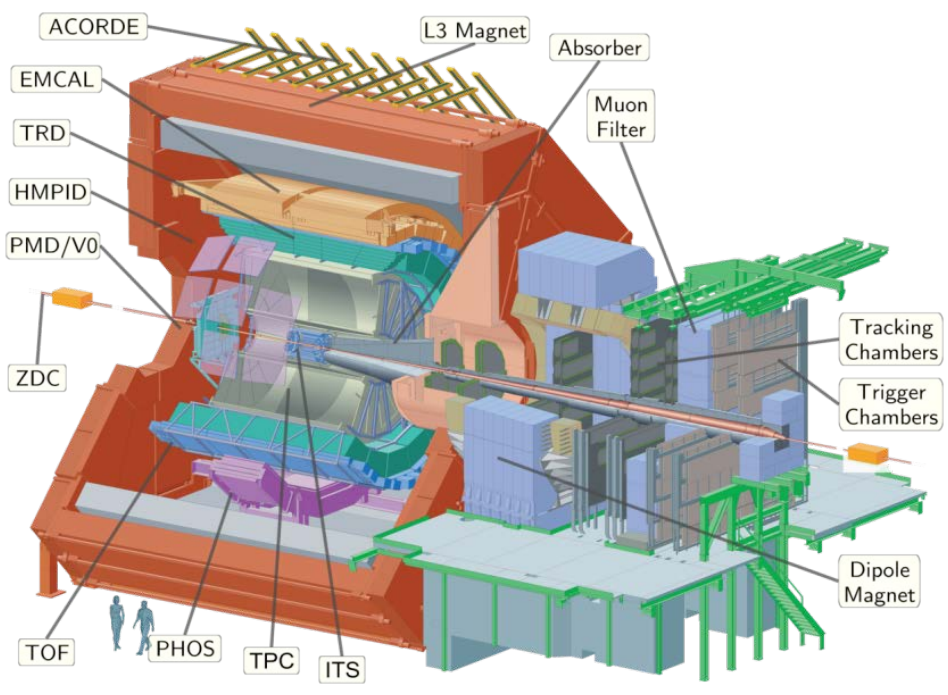}
\caption{The ALICE apparatus and its various detector components.}
\label{fig:alice}
\end{minipage}\hfill
\begin{minipage}{17.5pc}
\includegraphics[width=17.5pc]{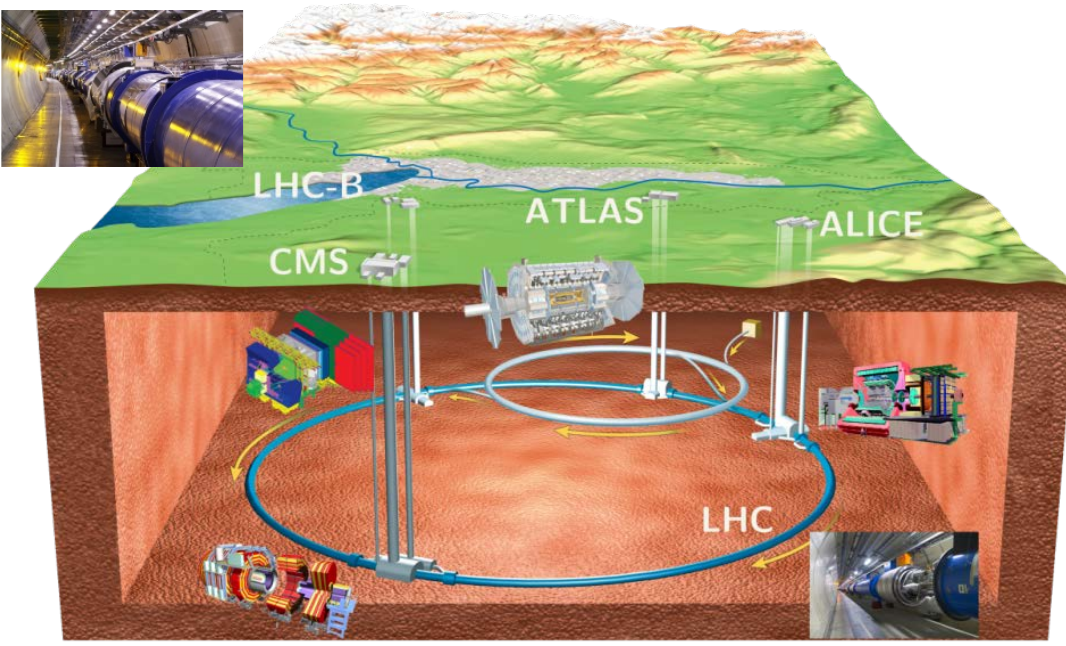}
\caption{The Large Hadron Collider beneath Geneva and its four major experiments.}
\label{fig:lhc}
\end{minipage}\hspace{2pc}
\end{figure}

ALICE (A Large Ion Collider Experiment~\cite{alice}, see Figure~\ref{fig:alice}) is one of four large-scale experiments at the LHC (Large Hadron Collider, see Figure~\ref{fig:lhc}) at CERN in Geneva.
Its main design purpose is the study of heavy ion collisions, which, compared to proton-proton collisions studied primarily by the other experiments, create a much higher number of particles in the collision.
ALICE employs a variety of detectors to identify the particles created in the initial collision and in secondary decays by measuring their trajectories and properties.
The main detectors for tracking the particle trajectories are the TPC (Time Projection Chamber) and ITS (Inner Tracking System).
Several detectors of ALICE are sensitive to environmental conditions such as ambient temperature and pressure and need proper calibration to deliver exact measurements.

\looseness=-1
The ALICE High Level Trigger (HLT) is an online computing farm of about 200 compute nodes that evaluate the data recorded by the ALICE detectors in real time.
The baseline is the event reconstruction, which recovers the physical properties from the signals received from the detectors.
The most compute intense task in this respect is track reconstruction.
Building on the reconstruction, the HLT performs a variety of other tasks.
This includes real time data compression that increases the number of collision events that can be stored on tapes permanently,
  real-time calibration of the detectors to improve the reconstruction,
  triggering for and tagging of physically interesting events for analysis,
  as well as monitoring and QA (Quality Assurance).

The HLT framework runs all the processing tasks in isolated components that communicate via a modern publisher-subscriber based message passing interface.
The framework needs to provide various features to enable all HLT tasks and to ensure stable operation guaranteeing high data taking efficiency.
New developments in the HLT for the LHC Run~2 have extended the list of required features.
In addition, the readout rate of the TPC, the main contributor to data volume, has doubled compared to Run~1 and the HLT computing farm has grown in capacity accordingly.
This poses additional challenges for the HLT framework.
Therefore, the HLT framework has been tuned and complemented with new features for LHC Run~2.
The most important challenges and requirements have been the following:
\begin{itemize}
 \item \textbf{Data rate:}
   One major concern is the total incoming data rate in terms of gigabytes per second that the HLT can handle.
   The rate has an upper bound defined by the number of incoming optical links and their link speeds.
   The HLT network, the output links to data acquisition (DAQ), and the processing capabilities must be dimensioned accordingly to process the data in real time.
 \item \textbf{Event rate:}
   The event rate does in fact not depend on the data rate linearly.
   Heavy ion events are much larger than pp collisions yielding a much higher data rate at the same event rate.
   Also the selection of detectors that are read out by the HLT has an influence.
   While the TPC is anyway limited to around~$2$\,kHz, there are fast detectors transmitting small data sizes at a high event rate.
   Thus, the same data rate can be recorded both with very low or very high event rate.
   The challenge in this regard is not the amount of data, but the merging of all the event fragments received on different optical links at a high rate.
 \item \textbf{CPU load:}
   The framework maintains plenty of internal lists for merging the event fragments, scheduling all processing components, and data buffer management and lifetime bookkeeping.
   In order to minimize the impact on and provide maximum compute resources to the actual event processing, the framework's CPU load should be minimized.
 \item \textbf{Configuration and startup time:}
   When ALICE starts a data taking run, many parameters influence the HLT operation, like the list of the detectors to read out, the current status of the ALICE magnets, beam and environmental conditions, as well as calibration and configuration settings.
   The startup process is split in two phases: the configure phase performs all preparations that are possible before stable beam is declared, while the engage phase starts the actual processes providing final settings available during stable beam.
   However, change of some settings like the list of detectors to read out might require a new configuration step also during stable beam.
   In order to maximize the data taking efficiency, these phases must be as short as possible, while the engage phase is more important than the configure phase.
 \item \textbf{Asynchronous transporting and failure resiliency:}
   One new feature commissioned at the beginning of Run~2 is online calibration of the ALICE TPC.
   In order to match online and offline calibration results, we employ the same source code by wrapping the offline reconstruction inside an HLT component.
   The calibration involves long-running tasks, however, not for every event.
   Running them in the event-synchronous HLT chain could stall the pipeline and fill up all the HLT buffers finally leading to back-pressure to the detectors pausing the read out.
   Thus, asynchronous processing is required.
   Since the calibration involves many lines of codes, which are also not under the control of the HLT, we must ensure that even fatal errors in the calibration code do not interfere with the normal HLT processing chain.
 \item \textbf{Zero-MQ message transport:}
   The components in the HLT framework are arranged in a graph, where the components are the nodes and the subscriptions are the borders.
   Because the processing is event-synchronous, the graph must be loop free.
   This is by design incompatible with the online calibration, where the calibration result created at the end of the chain needs to be fed back to the reconstruction at the beginning of the chain at a later time.
   We facilitate this by complementing the synchronous HLT transport with another asynchronous transfer scheme based on Zero-MQ.
   This Zero-MQ transport is also used as an interface for new QA components, for the online event display, to send QA histograms to DQM, and it is a prototype for the interface of the ALICE online / offline computing upgrade for Run~3.
   This new feature is described in~\cite{zmq}.
\end{itemize}

\section{Framework processing optimizations}

For stress-testing the HLT framework with high data and event rates and measuring CPU load in a controlled environment, we use data-replay.
In this mode, the FPGA-based C-RORC card (see~\cite{crorc}), which normally receives the data from the detectors, stores data from a set of events in its internal memory.
It can replay these events at a defined rate in an endless loop, which looks to the HLT exactly as if the events were coming from the detector.
By selecting events of certain sizes and with contribution from certain detectors, one can define the desired data and event rates.
In more detail, we use a selection of large events with TPC contribution to modulate the data rate, and add small events without TPC content to select the event rate.

An upper bound for the maximum rates deliverable from the experiment can be computed from the technical aspects.
The data rate is mostly defined by the TPC, which has~$216$ optical links at~$3.125$\,GBit/s, or~$84$\,GB/s in total.
However, the TPC cannot drive all links at maximum capacity at the same time because the number of TPC pad rows and the occupancy in the inner and outer TPC sectors differ.
Running in a condition that uses the full~$3.125$\,GBit/s for links with less data would exceed the link capacity for other links.
Taking this into account and subtracting the protocol overhead and 8b/10b encoding, the TPC is bound by~$48$\,GB/s.
On top of that, the TPC pauses the readout during the sampling to avoid side effects on the analog digital conversion such that the link cannot send data continuously.
This reduces the real readout rate further, depending on the event size.
Still, we assume~$48$\,GB/s for our tests giving us some margin compared to real operation.
The contribution from other detectors is small, for the events we selected it is below~$4$\,GB/s in total.

In the current HLT configuration, these~$52$\,GB/s correspond to a maximum of~$1.377$\,GB/s transferred over the network by an HLT input node.\footnote{This rate depends on a couple of points: detector, data rate of the detector, distribution of the data over the detector links, and in case of the TPC compression factor of the FPGA based TPC cluster finder.
The values stated here are the measured peaks during data replay.}
After processing the data and compression in the HLT, the maximum data rate received by an HLT output node over network is~$1.53$\,GB/s, and the maximum aggregate rate sent to data acquisition is~$10.7$\,GB/s.
We conducted measurements of the HLT framework's network performance by injecting extra data into the events up to a send and receive rate of~$2.4$\,GB/s per node, leaving a comfortable margin compared to the required rates.
The maximum data rate measured outwards to data acquisition is around~$12$\,GB/s matching more or less the aggregate link speed to DAQ of~$12.5$\,GB/s.
This covers the absolute peak requirement of our worst case scenario.
The rate to DAQ could only be increased by using additional fiber connections, which would be possible today using remaining currently unconnected HLT output ports.
The above scenario models the absolute worst case where all detectors send data at maximum link speed and there is still margin in the network infrastructure.
This proves that from a network transfer perspective, the HLT is able to handle every possible data taking scenario.

Before we show that the HLT is also capable of processing data at this rate, we discuss maximum event rates.
The TPC readout is limited to an absolute maximum of slightly below~$2$\,kHz, for large events it is much less.
Other detectors send events faster.
For the replay tests, we use pp and Pb-Pb data from different runs.
The largest events are from Run~245683, a period with the highest Pb-Pb luminosity and pile-up.
For this data set, the above data rate limit translates to a TPC event rate of~$950$\,Hz.
The bottleneck for the HLT is mostly the merging of the event fragments, in particular, the lists with all the present fragments and the inter-process communication of the framework processes responsible for receiving fragments, performing the actual merging, and shipping merged events to all subscribers.
The lists have already been optimized for Run~1 using hash tables and are fast enough also for Run~2 conditions without modification.
Inter-process communication, however, turned out to become a showstopper.
The HLT framework software after Run~1, but operating on the Run~2 farm, achieved a maximum rate of~$500$\,Hz in the worst case replay scenario, which is lacking a factor of two compared to the~$950$\,Hz.
Using smaller pp events and a stripped-down HLT configuration with several components disabled and only performing the baseline operation with TPC compression, a rate of~$3$\,kHz was achieved.
This is also not universally sufficient for all pp scenarios.
Consequently, we have improved the HLT framework for Run~2, also reducing the required CPU load.

In particular for high event rates, the old software required super-linearly more CPU cycles.
Inter-process communication was implemented via UNIX pipes requiring system calls for reading and writing.
We have replaced this approach by a shared memory based communication using spinlocks.\footnote{The processes are only exchanging metadata with pointers to buffers anyway, so actual read and write operations are fast and spinlocks are superior to operating system semaphores.}
In addition, we have implemented a mechanism to merge multiple messages into one communication where possible.
In particular, the RORC publisher process, which provides the events to the framework, can announce multiple events at once when they are small.
This allows a joint processing with single communication messages of all these events on the input node, where it is most critical.
In some cases, the human-readable configuration input file defined data relays to control the data-flow.
In case they are not needed, they are now optimized away, and the processes talk directly without relay.
Also, intervals for status polls have been increased, which led to significant load for the master task manager processes that were communicating with several thousands of processes.

\begin{table}
\centering
\caption{\label{tab1}CPU load of ALICE HLT framework processes before and after the tuning for Run~2.
Full utilization of one CPU core corresponds to~$100$\,$\%$ load.
Values are per input compute node.
The example is for a node with the maximum of twelve input links.}
\begin{tabular}{llrr}
\br
Data rate & $3$\,kHz & $6$\,kHz \\
Framework version & before tuning & after tuning \\
\mr
Framework process & CPU load & CPU load & \\
\mr
Event Merger & $240$\,$\%$ & $200$\,$\%$ \\
TaskManager & $100$\,$\%$ & $30$\,$\%$ \\
RORCPublisher & $12 \cdot 75$\,$\%$ & $12 \cdot 30$\,$\%$ \\
DataRelay & $80$\,$\%$ & $0$\,$\%$ \\
EventScatterer & $80$\,$\%$ & $60$\,$\%$ \\
\mr
Sum & $1200$\,$\%$ & $650$\,$\%$ \\
\br
\end{tabular}
\end{table}

\begin{figure}[b]
\includegraphics[width=0.67\textwidth]{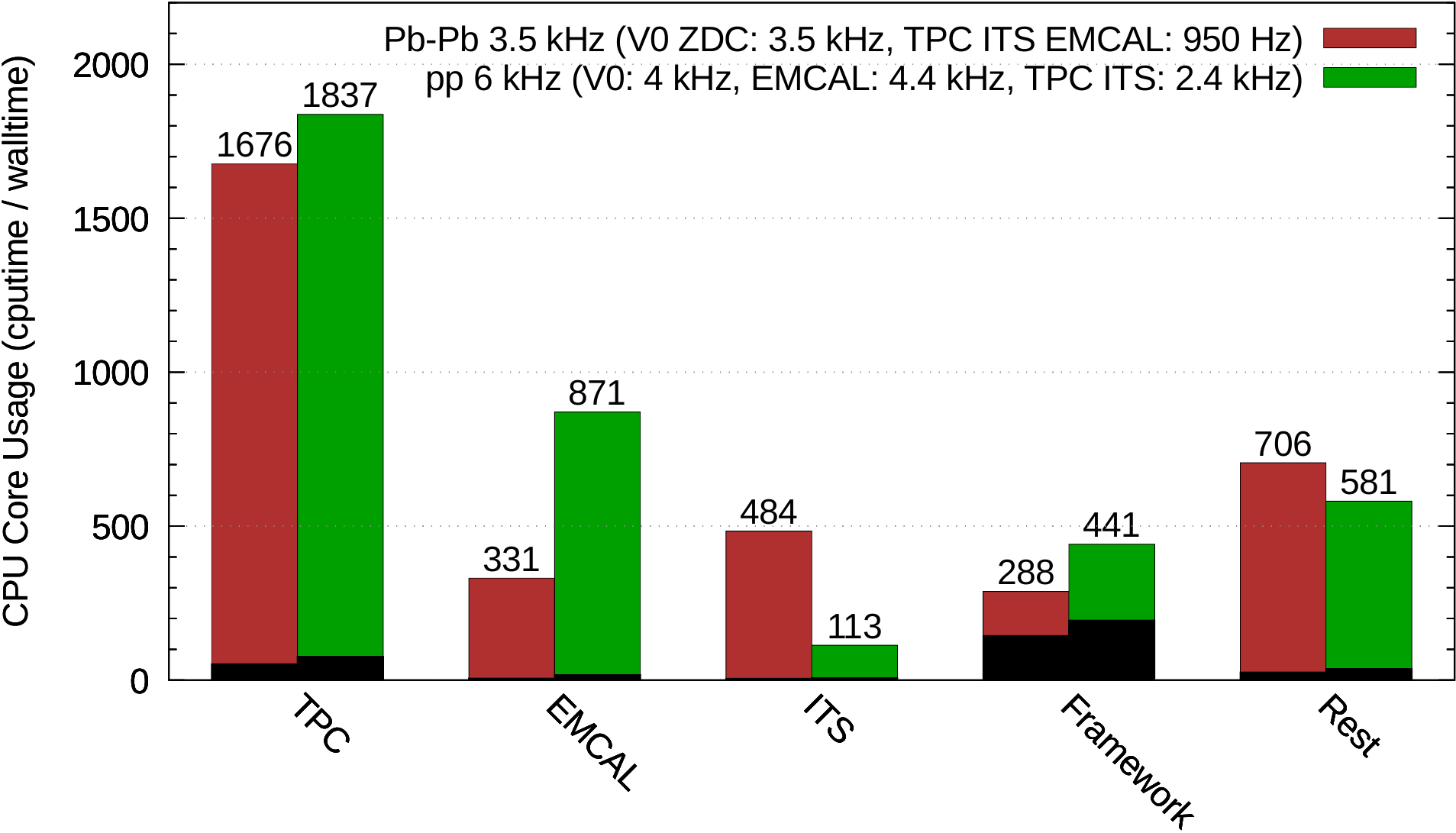}\hspace{2pc}%
\begin{minipage}[b]{0.28\textwidth}\caption{CPU utilization during HLT operation in Pb-Pb and pp data replay at maximum possible rate in terms of number of CPU cores (aggregate in the HLT cluster) used for several tasks.
System load of system calls (in contrast to user time) is shown in black.}
\label{fig:gp_cpu_util_sum}
\end{minipage}
\end{figure}

With all improvements, the mergers on the input nodes can now process 12 inputs (the maximum of the C-RORC~\cite{crorc}) at~$6$\,kHz, even with reduced CPU load.
Table~\ref{tab1} gives an overview of the CPU load of framework processes before and after the tuning for run 2.
Figure~\ref{fig:gp_cpu_util_sum} shows the CPU time used by the framework, and by reconstruction processes of several detectors.
It shows that the aggregate framework CPU load, including data input, network transfer, output, scheduling, and monitoring is between~$8$\,$\%$ and~$13$\,$\%$.
Roughly half the framework load is system load caused by the network transfer.
The TPC reconstruction uses GPU acceleration, and the GPU data transfer is visible as system load for the TPC~\cite{gpu}.

\begin{figure}[t]
\includegraphics[width=0.48\textwidth]{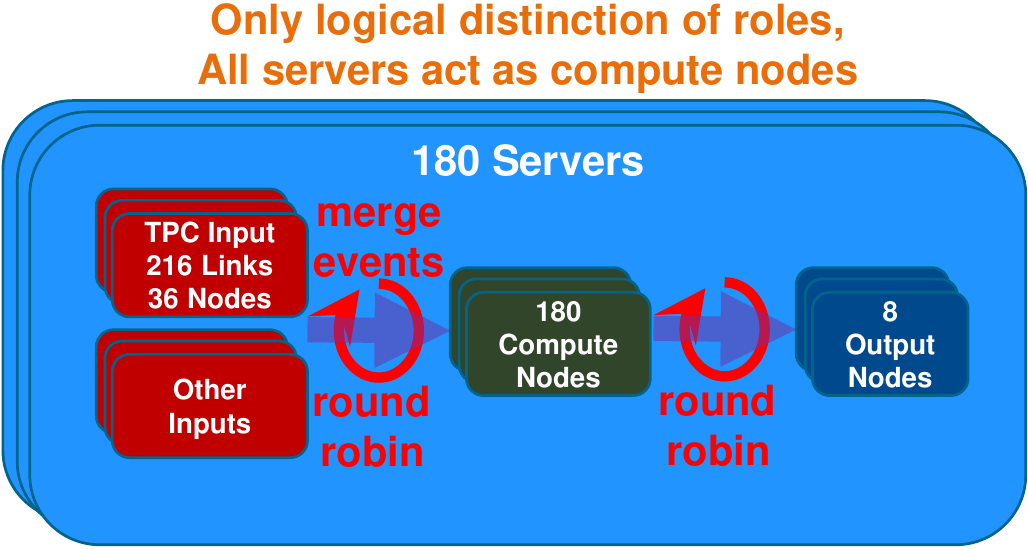}\hspace{2pc}%
\begin{minipage}[b]{0.50\textwidth}\caption{Illustration of the ALICE HLT processing approach.}
\label{fig:hltoverview}
\end{minipage}
\end{figure}

The reduction of the CPU load yields an additional benefit:
In contrast to the Run~1 farm, the Run~2 HLT farm is completely homogeneous with identical servers.
The only difference is that a subset of the servers is equipped with C-RORC cards to receive and send data via the DDL fibers.
The servers serve different roles and act logically as at least one of input node, output node, or compute node,
Prior to the framework improvements, the input nodes could not be used efficiently for the computation, because the input mergers and input components were causing excessive CPU load.
Now, the difference of the load between the input nodes and pure compute nodes is only roughly three CPU cores, such that we can employ a simple round-robin load balancing assigning every server the compute role without sacrificing much compute power.
Figure~\ref{fig:hltoverview} illustrates the HLT chain on a high level: the input nodes receive data from one to twelve DDL links and merge these fragments at the full event rate.
Then, they send the partially merged fragments in a round-robin fashion to all compute nodes (including themselves), where the full events are merged at a much lower rate.
The compute nodes (which means all physical servers) process a full event at once and then send it round-robin to the output nodes.

Tables~\ref{tab2} and~\ref{tab3} give an overview of data and event rates achievable in several scenarios.
The maximum possible experiment data rate is covered for both pp and Pb-Pb and the event mergers can process up to $6$\,kHz in any case.
The trigger scenarios for Run~2 foresee at maximum~$2$\,kHz minimum bias data taking for the central barrel detectors.
In addition, both the fast detector trigger group and the muon detectors can contribute additional some hundred Hertz each.
Thus, the HLT framework covers all ALICE data taking scenarios during Run~2 both from a data rate and event rate standpoint.
CPU processing performance is in all cases sufficient.
The average failure rate during Pb-Pb stress tests is around one in~$500$ million events or~$5$\,Pb of data.

\begin{table}
\centering
\caption{\label{tab2}Scenarios, participating detectors, and limiting factors of data replay tests.}
\begin{tabular}{lll}
\br
Scenario & Detectors & Limiting factor \\
\mr
Single input DDL & ZDC & Framework \\
pp $5.02$\,TeV Run 244364 & TPC, ITS, EMCAL, V0 & CPU load \\
pp $13$~TeV Run 239401 & TPC, ITS, EMCAL, V0 & DDL bandwidth \\
Pb-Pb Run 245683 & TPC, ITS, EMCAL, V0, ZDC & DDL bandwidth \\
Pb-Pb Run 245683 & ITS, EMCAL, V0, ZDC & Framework \\
\mr
\textbf{pp $\mathbf{13}$~TeV Run 239401} & All & DDL bandwidth / Framework \\
\textbf{Pb-Pb Run 245683} & All & DDL bandwidth / CPU \\
\br
\end{tabular}
\end{table}

\begin{table}
\centering
\caption{\label{tab3}Maximum data and event rates achieved in data replay scenarios listed in Table~\ref{tab2}.}
\begin{tabular}{lrrr}
\br
Scenario & Data rate & TPC event rate & Total event rate \\
\mr
Single input DDL & $6$\,MB/s & 0 & $10$\,kHz \\
pp $5.02$\,TeV Run 244364 & $8.3$\,GB/s & $4.5$\,kHz & $4.5$\,kHz \\
pp $13$~TeV Run 239401 & $48$\,GB/s & $2.4$\,kHz & $2.4$\,kHz \\
Pb-Pb Run 245683 & $48$\,GB/s & $950$\,Hz & $950$\,Hz \\
Pb-Pb Run 245683 & $3.5$\,GB/s & 0 & $6$\,kHz \\
\mr
\textbf{pp $\mathbf{13}$~TeV Run 239401} & $49$\,GB/s & $2.4$\,kHz & $6$\,kHz \\
\textbf{Pb-Pb Run 245683} & $51$\,GB/s & $950$\,Hz & $3.75$\,kHz \\
\br
\end{tabular}
\end{table}

\section{Improvements of startup time}

\begin{figure}[t]
\includegraphics[width=0.89\textwidth]{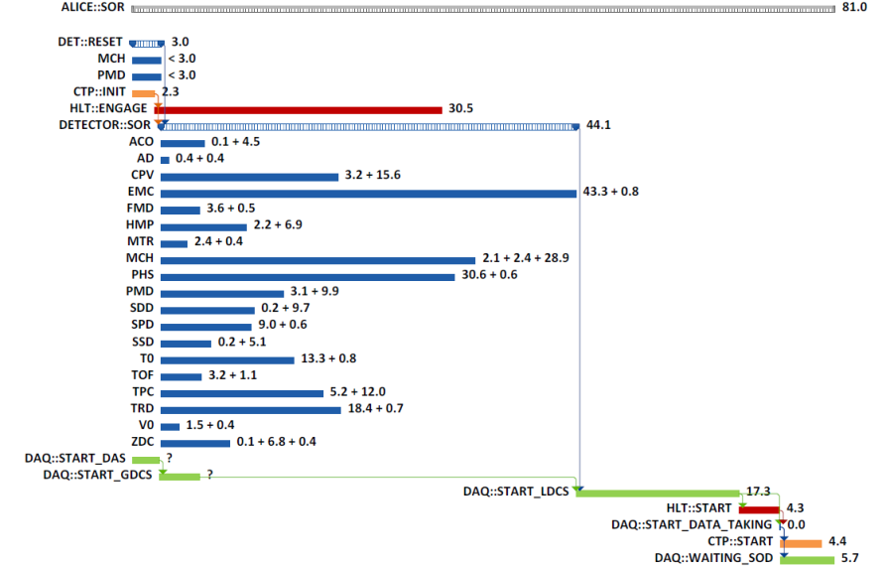}\hspace{2pc}%
\begin{minipage}[b]{1.0\textwidth}\caption{Visualization of the startup time of all ALICE subsystems - HLT highlighted in red. \textit{(From Vasco Barroso, ALICE Run~2 2015 closeout workshop)}}
\label{fig:startup}
\end{minipage}
\end{figure}

The goal for the HLT startup time is to be fast enough to remain in the shadow of the detector startup.
Figure~\ref{fig:startup} shows the situation for HLT engage time at the end of 2015.
With~$30.5$\,s the HLT was still faster than EMCAL, PHOS, and the Muon Chambers.
Since the detectors have been improving their startup time, HLT needed to accelerate its own as well.
With regard to the configure time, HLT was the slowest system beforehand already in 2015.
Adding additional HLT components for new features like online calibration further prolong the configure step, which would have been unacceptable without improvement.

The configuration phase consists of Python code to read a high level XML configuration, create the configuration file for all HLT processes, and export the configuration to an XML file.
Due to the Python global interpreter lock, multi-threading cannot speed up the script, while multiprocessing is unsuited for operation on the global configuration classes but it could be applied to the creation of the output files, which are independent from each other.
The remaining Python code was optimized manually, improving several of the algorithms for process placement, using bitmaps instead of hash-maps to accelerate set operations, and adding caches of intermediate results where possible.
We use the static Cython compiler for Python, and we run other configuration steps in parallel to hide them in the shadow of the Python script.
Overall, we improved the input phase from~$30$ to~$1.5$\,s, the processing phase from~$160$ to~$13$\,s, and the output phase from~$20$ to~$2$\,seconds for the maximum configuration with all processes.

Shortening the engage phase is more challenging with less code improvement margin.
Some improvement comes already from the improved inter-process communication.
The most important step is the movement of a part of the engage steps into the configure phase.
Because this approach leaves less parallelism for the different tasks, the aggregate time of configure and engage increases, but it brings the engage time down to~$17$\,s, while the configure step is already fast enough.
In order to estimate an upper bound for the startup time in case the HLT configuration grows further, we conducted a test with readout and data transport for all detectors in (even if unused), and many additional test components in the chain.
This resulted in~$27$\,s for configure and~$22$\,s for engage, which is still in the shadow of the detectors.
Therefore, with these improvements the HLT causes absolutely no delay anymore during startup of a run.

\section{New HLT framework features}

For running long calibration tasks in the event-synchronous HLT chain, we have added a feature to support asynchronous tasks~\cite{chep15}.
The framework spawns an additional thread during processing an event, passes the data, and then finishes the current event while the task is still running.
When the asynchronous task is finished, the result is integrated in the chain during the processing of a later event.
This is sufficient for the calibration, which anyway needs to accumulate statistics over many events, and can then be used for a certain amount of time.

One problem we have in ensuring stability is that the offline calibration code is not under HLT control.
With the original threading approach, a segmentation fault in the calibration process would kill the entire process, including the framework part.
We have thus complemented the asynchronous feature with the possibility to spawn a process instead of a thread and pass data via shared memory.
This imposes a small penalty for data copies that might be needed, but it guarantees that the HLT keeps running when the calibration encounters a fatal error.
In that case, the asynchronous process can simply be restarted.

\section{Conclusions}

\begin{figure}[htb]
\includegraphics[width=0.93\textwidth]{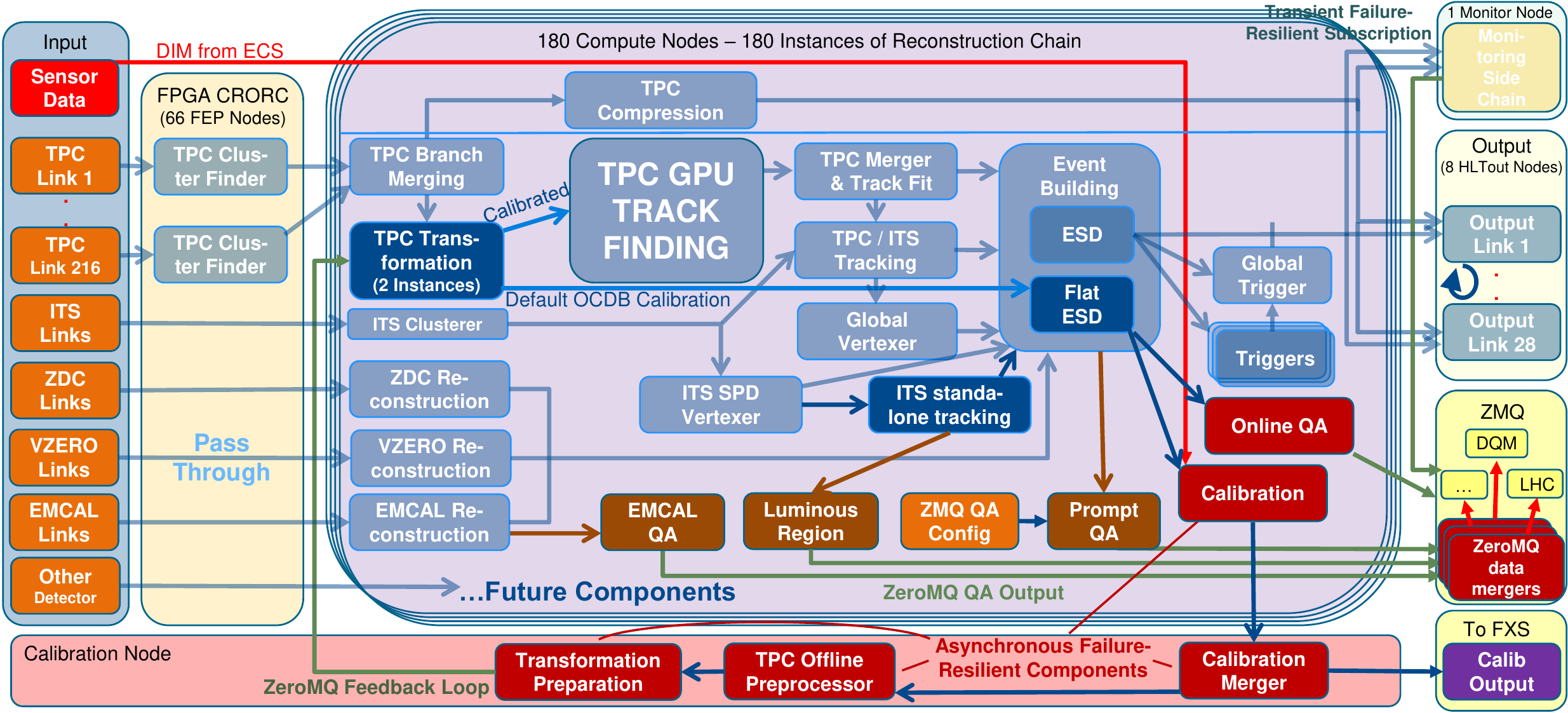}\hspace{2pc}%
\begin{minipage}[b]{1.0\textwidth}\caption{Simplified illustration of all components currently running in the HLT.
Components with light background have been running in since 2015, while components with dark background have been added in 2016 thanks to the framework improvements.}
\label{fig:hltchain}
\end{minipage}
\end{figure}

The ALICE HLT framework has been tuned to cope with higher data and event rates with the TPC readout upgrade starting in 2016.
It was proven that it supports all possible ALICE data taking scenarios.
The startup procedure was tuned to shorten the start up time improving data taking efficiency.
New features were added to facilitate additional tasks like TPC online calibration.
Figure~\ref{fig:hltchain} gives a simplified overview of processing components running in the HLT in 2016, highlighting new components commissioned in 2016 with dark background colors.

\end{document}